\documentclass[12pt]{article}
\usepackage{esfconf}
\usepackage{epsfig}

\newcommand{\beq}{\begin{equation}}
\newcommand{\eeq}{\end{equation}}
\begin{document}


\title{Brane tunneling and the Brane World scenario}


\authors{A.Gorsky, K.Selivanov}


\addresses{ ITEP, B.Cheremushkinskaya, 25, 117218, Moscow}


\maketitle


\begin{abstract}
Fate of branes in external fields is reviewed. Spontaneous creation
of the Brane World in $AdS_{5}$ with external field is described.
The resulting Brane World consists of a flat 4d spatially finite expanding
Universe and curved expanding "regulator" branes. All branes have a positive
tension.
\end{abstract}


During the last years branes where recognized
as the important ingredient of the
complete picture  in the string and field theories.
The branes actually have the dual nature; they can be considered as
the fundamental objects where the open strings can end on
and as the solitons in the supergravity theories.
The branes below can be viewed on as the fundamental
ones or as effective ones, e.g. domain walls.
Domain walls typically appear in the context
of N=1 SUSY gauge theories and since they saturate BPS
condition they can be treated quasiclassically.
In what follows we shall focus on the behaviour of the branes
in the external fields and the applications of some
quasiclassical processes involving branes to the
Brane World scenario.

The low-energy brane  action consists of two pieces, a tension term and
a charge term.  Tension term reads
\begin{equation}
\label{tension2}
S_{tension}=T \oint \sqrt {det({\hat g_{ind}})}
\end{equation}
where the
integration is over the brane world-volume, $g_{ind}$ is the metric induced
on the world-volume via its embedding into the target space and
$T$ is the tension of the brane.  This term is an analogue of the
mass term for a particle
\begin{equation}
\label{tension1}
m \oint
\sqrt{det({\hat g_{ind}})}
\end{equation}
where the integration is over
the particle world-line.\\
The charge term looks as follows
\begin{equation}
\label{charge1}
S_{charge}=e\oint {B},
\end{equation}
where ${B}$ is a $n$-form gauge field field,
corresponding curvature being $H=dB$.
Branes are the sources for this field and they are affected
by this field.
In what follows we assume $H$ to be a top degree form.

Everything we are  going to discuss here is related to the
Schwinger type process - production of branes by homogeneous
external field $H$.

\section{A warm up example}
Consider first a warm up example: production of the particles by a
homogeneous $E$ field.

For this case the effective action reads
\beq
\label{action}
S_{eff}=TL-eEA
\eeq
where $L$ is the length of the  world-line of the particles produced,
$T$ is a mass  of the particle,
(as usual, particle-antiparticle history looks like  a closed world-line),
and  $A$ is the area surrounded by the world-line.

Notice, by the way, that
upon  the appropriate identification
of the parameters, the same effective action describes
false vacuum decay in (1+1) scalar field theory.  Particles are
substituted by kinks, electric field - by the energy difference between
false and true vacua.

Extremal world-line for the action Eq.(\ref{action}) is a circle
of the radius
${\bar r}=\frac{T}{eE}$
therefore the probability P of the spontaneous process looks as
follows
\beq
P \propto exp(-\frac{\pi T^2}{eE}).
\eeq
Notice that Minkowskian evolution is obtained from the Euclidean bounce
by the analitical continuation.

\section{Spontaneous production of branes}
The above warm-up example is easily generalized to the case
of the spontaneous production of branes in a homogeneous $H$ field
\cite{bt}. The effective action
in the assumption of constant $H$ reads
\beq
\label{actione}
S_{eff}=TA-ehV.
\eeq
World-volume of the brane  produced forms a closed hypersurface
(like the world-line of the particle produced in the above example).
$A$ in Eq.(\ref{actione}) is the area of the world-volume,
$V$ - volume of the region inside the brane.
The  Euclidean bounce is a $p+1$-dimensional sphere of radius
$\bar{r}=\frac{(d-1)T}{eh}$.
The value of the effective action on the bounce yields the probability
of the brane production:
\beq
P \propto exp(-const \frac{T^{d}}{(eh)^{d-1}}).
\eeq

Analogously to the above example, The Minkowskian evolution can be obtained
from the Euclidean bounce via the analytical continuation.

\section{Induced brane production}
Let us now consider the induced brane production \cite{gs1}.
To this aim we have
to introduce a new ingredient - brane junctions
\cite{john}. Brane world-volumes can meet along the junction manifold.
Angles at which
the branes meet each other are fixed by the tension force balance
condition. Of course at the junction the charge conservation condition   is
fulfilled. The fact of the key importance is that junctions are
BPS configurations therefore they are classically stable and can
be used in the quasiclassical considerations. The BPS property of the
junction of the fundamental branes is more or less clear while
the similar argument for the effective branes follows from the
existence of the specific central charges in SUSY algebra. These
central charges were found in WZ model in D=4 \cite{wz} as  well
as N=1 SYM theory \cite{gs}. The effective junctions have been
also recognized in the M theory framework \cite{ga}.

The setup for the induced brane production is as follows. There is
external homogeneous field and there is external neutral brane which can
have junctions with the charged branes to be produced. The question is
what is the probability of such process.

Let us again begin with a warm up example - one particle induced decay in
(1+1) \cite{sv}.  One has a massive particle in the false vacuum
and the correspondig field has a zero mode on the wall which
could separate false and true vacua. Then, if a bubble of the true vacuum
inside the false vacuum is produced, it is more profitable for the
particle to ride a part of its way in the form of zero mode on the wall
of the bubble. The effective action describing this case,
\beq
\label{actioni}
S_{eff}=TL-eEA + m {\tilde L},
\eeq
differs from Eq.(\ref{action}) only by the last term where $m$ is the mass of
the extra particle, ${\tilde L}$ is the length of the extra particle
world-line (only outside the bubble).

\begin{center}
\epsffile{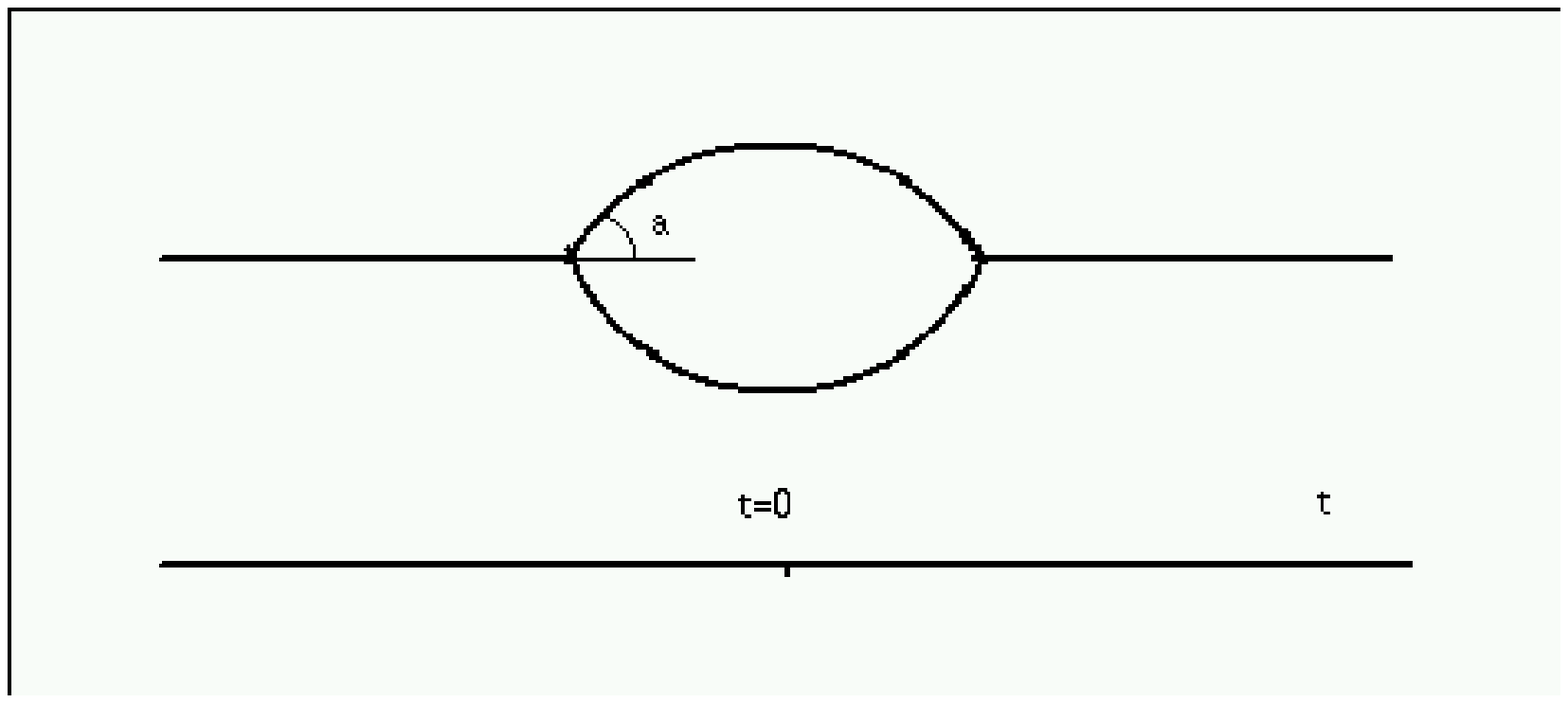}
\end{center}

The bounce is now glued from the  two segments of a circle of the same radius
as in the case of the spontaneous decay. These two segments meet  with the
world-line of the particle (junction) at the angle $\alpha$ which is defined
by the force balance condition, $m=2T \cos \alpha$.

The "charge conservation" in the present case is equivalent to a trivial fact
that when one passes through the bubble crossing its wall twice, one gets
back to false vacuum.
One then straightforwardly compute the probability of the false vacuum decay
\cite{sv}. There are two clear limiting cases. When mass of the particle is
small
compared to the mass of the wall, bounce is not disturted and
the probability is the
same as in spontaneous case. When mass of
the particle is close to $2T$, bounce shrinks
to a point and there is no exponential suppression in the induced vacuum decay.

Generalization to the case of branes has been developed in \cite{gs1}.

\begin{center}
\epsffile{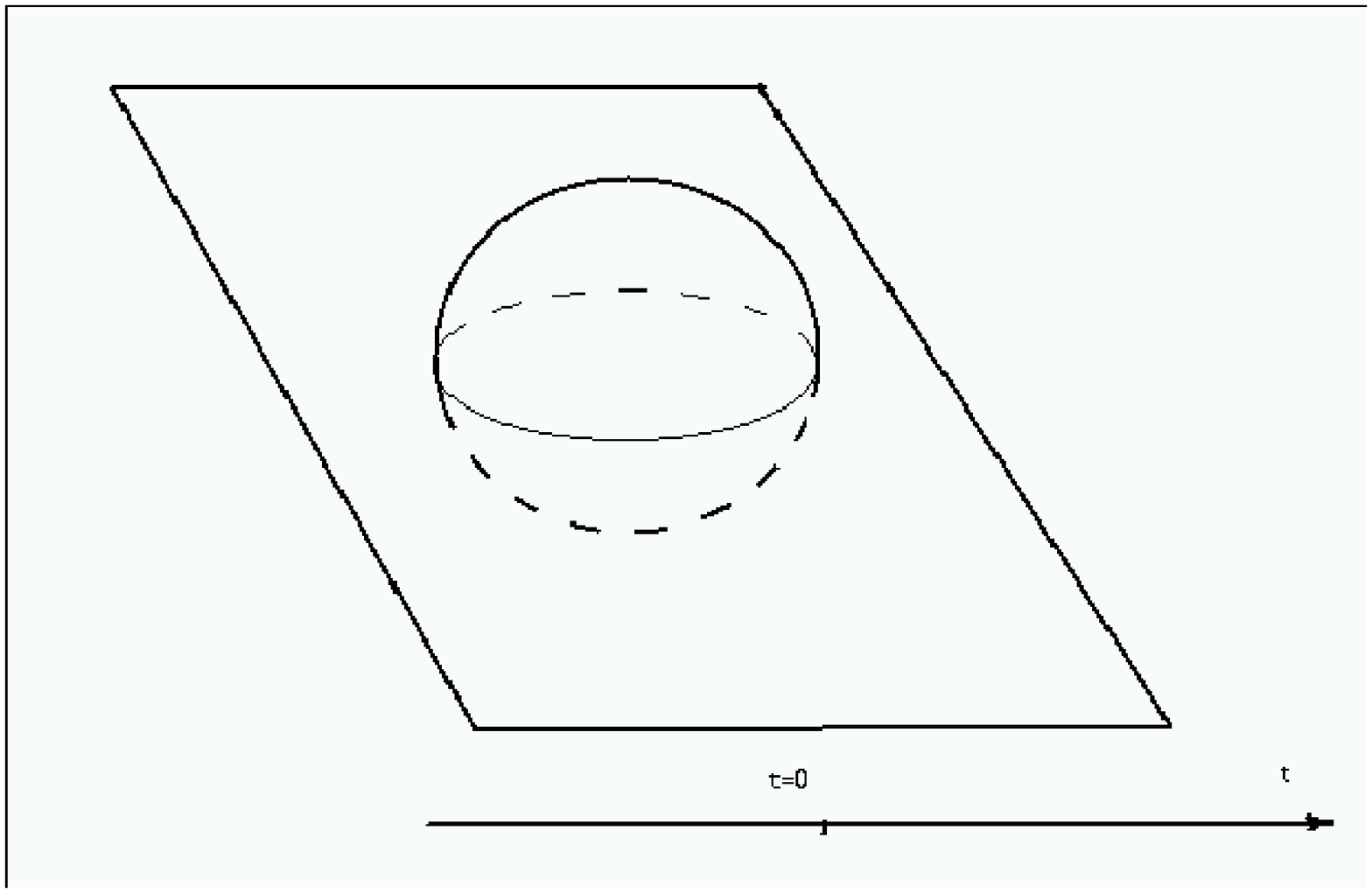}
\end{center}

The bounce consists now from two segments of (d-1) dimensional sphere of the
same radius as in the case of the spontaneous brane production,
glued to the external brane along the junction manifold. The force balance
condition reads
${\tilde T}=2T \cos \alpha$,
where ${\tilde T}$ is the tension of the external neutral brane.
Minkowskian evolution follows from the analytical continuation of the
bounce.
The calculation of the probability is straightforward and two clear limiting
cases are the same as in the particle case.

Let us mention the interesting application of such process \cite{gs1}. Let us
consider the noncommutative monopoles that is monopole solutions
when the external B field is switched on. It is known that
monopoles can be represented by D1 string ending on D3 branes \cite{dia}.
In the noncommutative case one can consider the exact solution to the
classical equations of motion in U(1) theory \cite{gn}. If there is
a constant $H=dB$ then according to the
process above the D1 string is exponentially unstable with
respect to the decay into the dyonic strings. From the point
of view of U(1) theory on D3 worlvolume it means that
monopole in the case of nonconstant noncommutativity
nonperturbatively decays into dyons.

\section{A sketch of Brane World}
Now let us make a digression to sketch an idea of the Brane World.
This is an alternative to the idea of compactification. In
the compactification approach
extra dimensions are compact and small and thus
cannot be seen at the moderate energies.
In the Brane World, extra dimensions are infinite, but the matter
\cite{rubshap}, gauge fields \cite{pw} and gravity \cite{rs} are localized on
a brane (domain wall or other topological defect in the extra dimensions).

The model considered in \cite{rs} involves two branes localized at different
points on a circle (the 5th dimension is
$S^{1}$ of arbitrarily large radius).
One of the branes provides the physical
worldvolume for d=4 (RS-brane), the other is a so-called
regulator brane (R-brane). The gravity is
localized on the RS brane. The drawback
of the model is that R-brane has a negative tension (see e.g. discussion in
\cite{witten,kl}).

Many modifications of the construction in \cite{rs} were studied
(see e.g. \cite{more} and other references to the original paper
\cite{rs}), most of them includes the negative tension brane.

Let us consider in more detail the construction in \cite{grs}.

\begin{center}
\epsffile{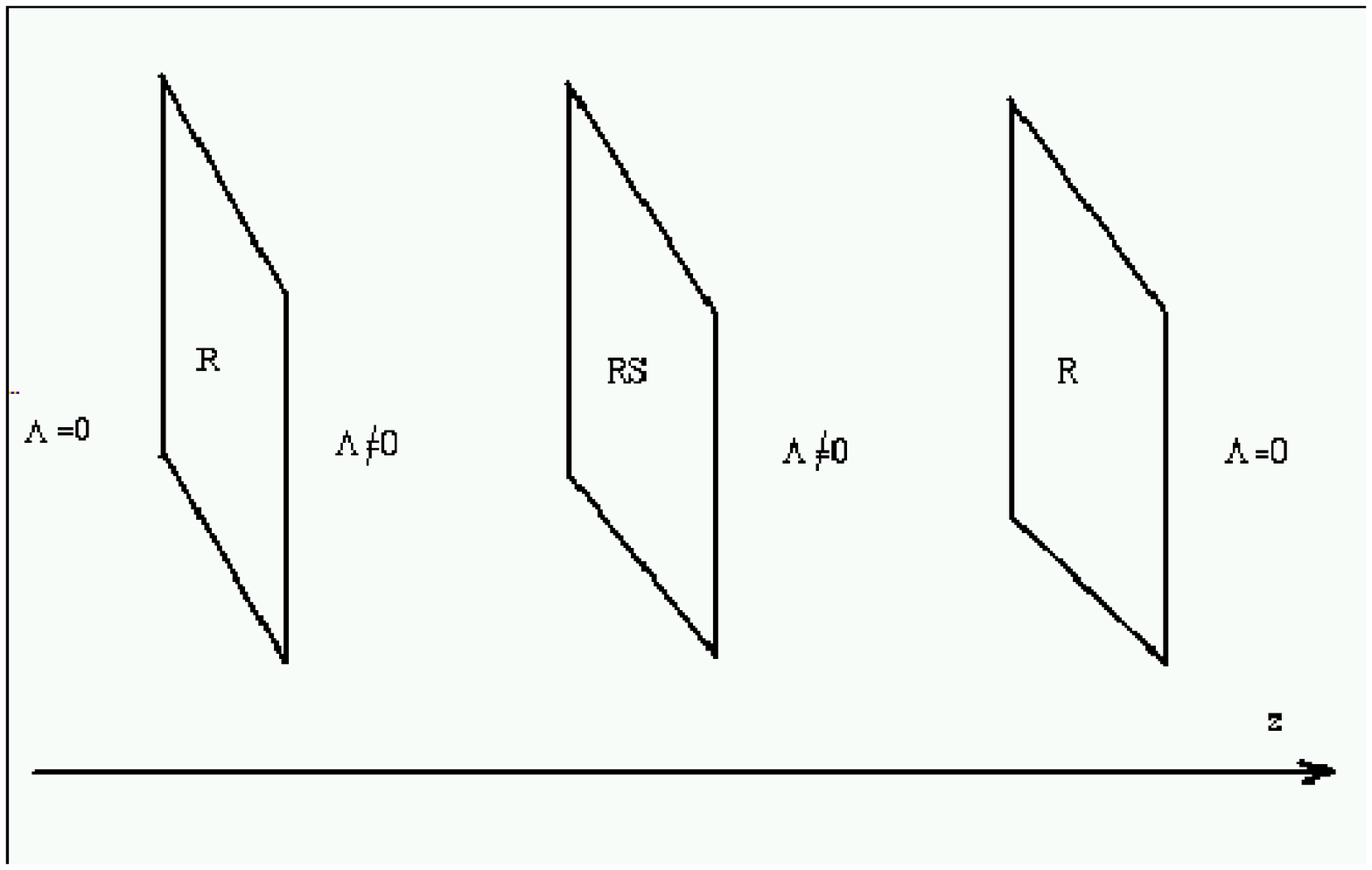}
\end{center}

It includes three branes localized on a line (the 5th dimension is
$R^{1}$), two negative tension R-branes and one RS-brane between them.
Cosmological constant outside R-branes is zero, cosmological constant
between R-branes and RS-brane is negative. Notice that in this model
cosmological constant is not really a constant, it jumps on the R-branes. So,
in fact, there is a hidden external field in the model, and R-branes are
charged with respect to it. This motivates the following construction in
\cite{gs2}.

\section{Spontaneous creation of the Brane World}
In \cite{gs2} a spontaneous creation of the Brane world
in a homogeneous external field was described. The process considered is a sort
of inverse
to the induced brane production in the external field.
Its particle counterpart is the spontaneous creation of the
charged pair with the additional neutral particle in the electric field.
The calculation of its
probability $ P \propto exp(-S)$ is straightforward and yields
\beq
S=\frac{2T_{R}^2}{eE}(\pi - arcsin(1- \frac{T_{RS}^2}{4T_{R}^2})^{1/2}) +
\frac{2T_{R} T_{RS}}{eE}(1- \frac{T_{RS}^2}{4T_{R}^2})^{1/2}
\eeq

The bounce now consists
of two segments of the charged branes (R-branes) which are glued along the junction
manifold with the neutral brane (RS-brane) which is inside the bubble.

\begin{center}
\epsffile{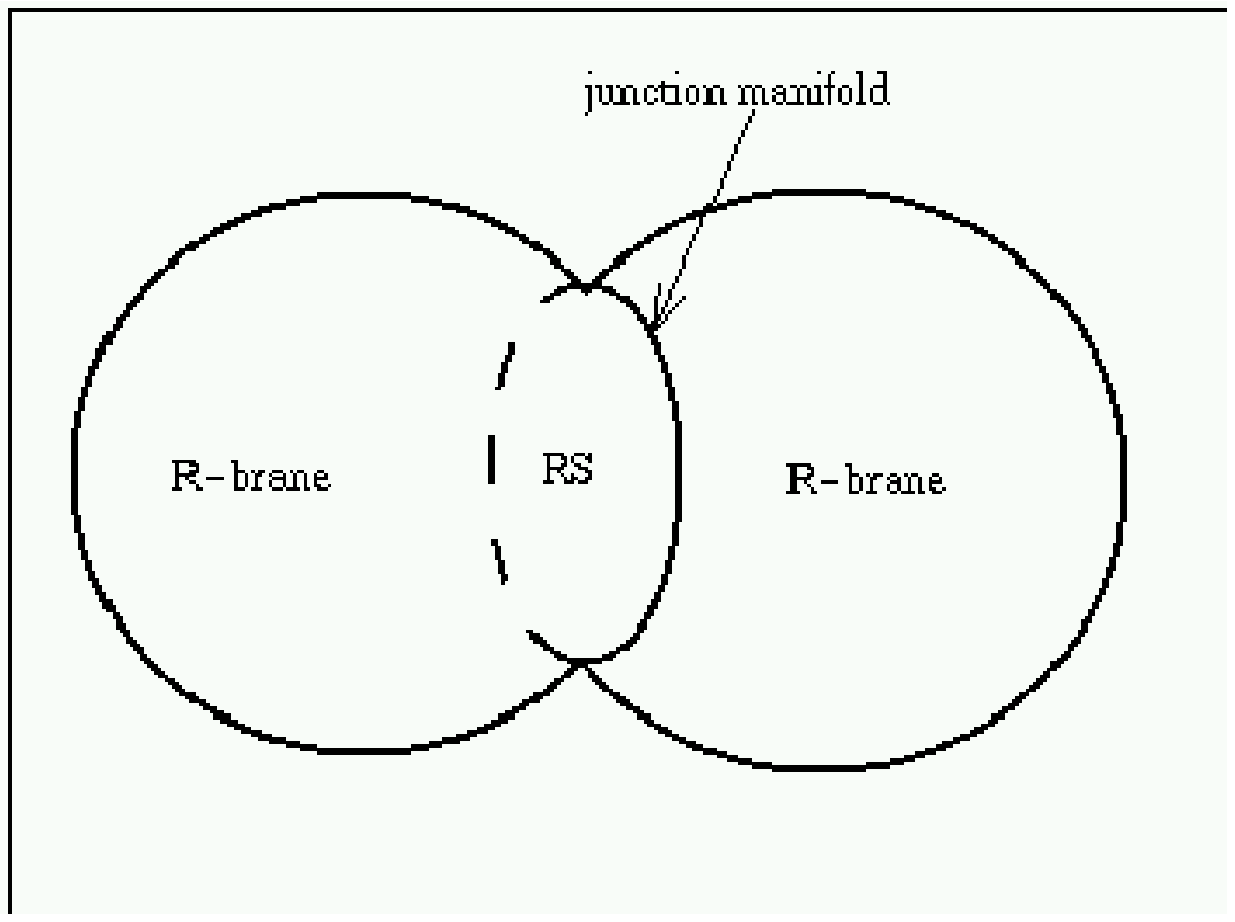}
\end{center}

Remarkably,  Einstein equations can be explicitely solved for
the quite complicated  brane configuration.
The metric consists of two AdS pieces sewed with each other
on the RS-brane  and with outside flat metric on the R-branes.
RS-brane inherits AdS metric, which however can be done as flat
as one wishes.
Of course, the metric of the "Euclidean" classical solution
determines
the ``Minkowskian'' evolution of the Universe. Importantly,
the R-branes are not static, they accelerate as they should
in the external field.  RS brane remains at rest.

After continuation into Minkowskian space our Universe is a spacially finite,
growing slice of AdS space. The spacial radius of R-brane is also growing,
and we
are eventually left with the picture of \cite{rs}, or \cite{grs}, type.
A peculiar property of the process described is that RS-brane is not flat, it
inherits $AdS_{4}$ metric so 4d cosmological constant problem remains unsolved.
To cure this peculiarity, in \cite{gs3} we have described the spontaneous
creation of the Brane World with flat RS-brane.

\section{Big Bang in $AdS_{5}$ with external field}

The setup in \cite{gs3} is as follows. There is $AdS_{5}$ with
nonvanishing homogeneous top degree form field and charged test branes.
The aim is to construct a tunneling into the Brane World
such that the resulting RS-brane is a peace of the same
flat section  of $AdS_{5}$
as in \cite{rs}.
Thus our 4d Universe living on RS-brane is flat and spatially finite
and we can take for granted from
\cite{rs} the 4d localization of gravity on the RS-brane and
the validity of 4d Newton law far enough in the future.
RS-brane is restricted by junction manifold where RS-brane meets with
the rapidly expanding R-branes.

Brane production in this setup has been studied in \cite{mms}.
The study essentially
is reduced to the consideration of the minimal charged surfaces in AdS with
the homogeneous field. These surfaces were (up to minor
subtleties) described in \cite{mms}. They are classified into three
classes: undercharged ones (saturating a sort of BPS inequality
between charge and tension of the brane),
overcharged ones (breaking the BPS inequality) and BPS ones .
Actually, only overcharged ones were in \cite{mms} given a tunneling
interpretation, the others have infinite volume. Our innovation compared
to \cite{mms} is to include junctions of those surfaces.
Our bounce now is glued out of three peaces of branes
- of BPS one, of undercharged one and of overcharged one (see fig.5). The BPS
brane plays the role of RS-brane, the others are R-branes.
Notice that the overcharged branes inherit the
metric of the sphere, the undercharged branes
- the one of AdS, and the BPS brane - the flat one, we shall
discuss this point in more details later.

\section{Charged surfaces in AdS}
We now describe, essentially
following \cite{mms}, "Euclidean"
minimal charged surfaces, that is,  solutions of test brane worldsheet
equations relevant for tunneling.

Let us consider the metric of AdS as in \cite{mms},
\beq
\label{metric}
ds^2 =  R_{ads}^2 \left(\cosh^2\!\rho \, d\tau^2 + d\rho^2 + \sinh^2\!{\rho} \,
d\Omega_{d-2}^2 \right),
\eeq
where $R_{ads}$ is the anti-de Sitter radius, and assume that
the curvature form $H=dB$ of the $B$-field is proportional to the volume form
with a constant coefficient (flux density). Assuming
the spherical symmetry
of the brane worldsheet, one reduces the effective action
to :
\beq
\label{action2}
S = TR_{ads}^{d-1} \Omega_{d-2}  \int d\tau \left[
\sinh^{d-2}\rho  \sqrt{ \cosh^2\rho + \left({d \rho
\over d \tau}\right)^2 } - q \sinh^{d-1}\rho  \right]
\eeq
where
$\Omega_{d-2}={2 \pi^{{(d-1)\over 2}} \over \Gamma({d-1 \over 2})}$
stands for the  volume of a unit d-2 sphere
and  $q$ is a constant made out of the flux density, brane charge
and the brane tension. The condition $q=1$ is identified in \cite{mms} as the
BPS one.  The branes with $q<1$ will be referred to as undercharged ones and
those with $q>1$ - as overcharged.

Before describing them we would like
to relate the metric Eq.(\ref{metric}) used in \cite{mms} to more
canonical ones. Upon the change of variables,
\beq
\label{variables2}
\tanh \tau = \tan \theta
\eeq
the metric  Eq.(\ref{metric}) is put to the form
\beq
\label{metric2}
ds^2 =  R_{ads}^2 \frac{d\tau^2 + d\theta^{2} + \sin^{2}\theta \,
d\Omega_{d-2}^2 }{\cos^{2}\theta}
\eeq
Finally, upon the change of variables
\beq
\label{variables3}
z=e^{\tau} \cos\theta,\;r=e^{\tau} \sin\theta
\eeq
one obtains the following canonical form of AdS metric
\beq
\label{metric3}
ds^2 =  R_{ads}^2 \frac{dz^2 + dr^{2} + r^{2}\,
d\Omega_{d-2}^2 }{z^{2}}.
\eeq

The minimal surfaces for the case  $q<1$
look as follows:
\beq
\label{under}
\cosh\rho = { \sinh \tau_{m} \over \sinh (\tau +a) }
\eeq
where $\tanh \tau_{m}=q$.

In coordinates Eq(\ref{variables2}) the undercharged surfaces
take the form
\beq
\label{under2}
\cos\theta=\frac{\sinh(\tau + a)}{\sinh\tau_{m}}.
\eeq
Restriction of the $AdS_{5}$
metric Eq.(\ref{metric2}) onto the undercharged surfaces Eq.(\ref{over2})
gives the metric of $AdS_{4}$.

In the case  $q>1$
the relevant charged minimal surfaces look as follows:
\beq
\label{over}
\cosh\rho = { \cosh \rho_{m}\over \cosh (\tau +b) }
\eeq

where  $\tanh \rho_{m} = 1/q$.
In coordinates Eq(\ref{variables2}) the overcharged surfaces
take the form
\beq
\label{over2}
\cos\theta=\frac{\cosh(\tau + b)}{\cosh\rho_{m}}.
\eeq
Restriction of the $AdS_{5}$
metric Eq.(\ref{metric2}) onto the overcharged surfaces Eq.(\ref{over2})
gives the metric of the sphere $S_{4}$.
The BPS case (q=1) can be obtained as a limit from either of the cases
above. Corresponding surfaces look as follows :

\beq
\label{bps}
\cos\theta=\frac{1}{z_{0}}e^{\tau}
\eeq
where $z_{0}$ is a constant.
Upon the inversion transformation one obtains
\beq
\label{bps2}
e^{\tau} \cos\theta=z_{0}
\eeq
In terms of Eq.(\ref{metric3}) these are the surfaces $z=z_{0}$ .
Apparently, restriction of the AdS metric onto these surfaces
gives flat Euclidean metric.

Only overcharged surfaces admit a tunneling
interpretation \cite{mms},
since undercharged and BPS ones reach the boundary of AdS
space and thus have infinite volume and infinite effective action.\\

\section{The solution}
We shall now construct the bounce which describes the tunneling into the
Brane World \cite{gs3}. It is glued out of three pieces - a piece of BPS brane
located along $z=z_{0}$ section, Eq.(\ref{bps2}), and playing a role
of RS-brane in the
Brane World, a piece of undercharged brane, Eq.(\ref{under2}),
located above the
BPS brane (in the fixed coordinates of the type of Eq.(\ref{metric3})
and playing a role of one R-brane, and a piece of overcharged brane,
Eq.(\ref{over2}),
located below the BPS brane and playing the role of the other R-brane
(see fig.5). All three pieces are glued along the junction manifold.
The usual junction conditions are the charge conservation and
the tension forces balance at the junction \cite{john} .
We shall assume that there is no junction energy contribution
to the effective action,
though this assumption is not crucial for the existence of solution.

\begin{center}
\epsffile{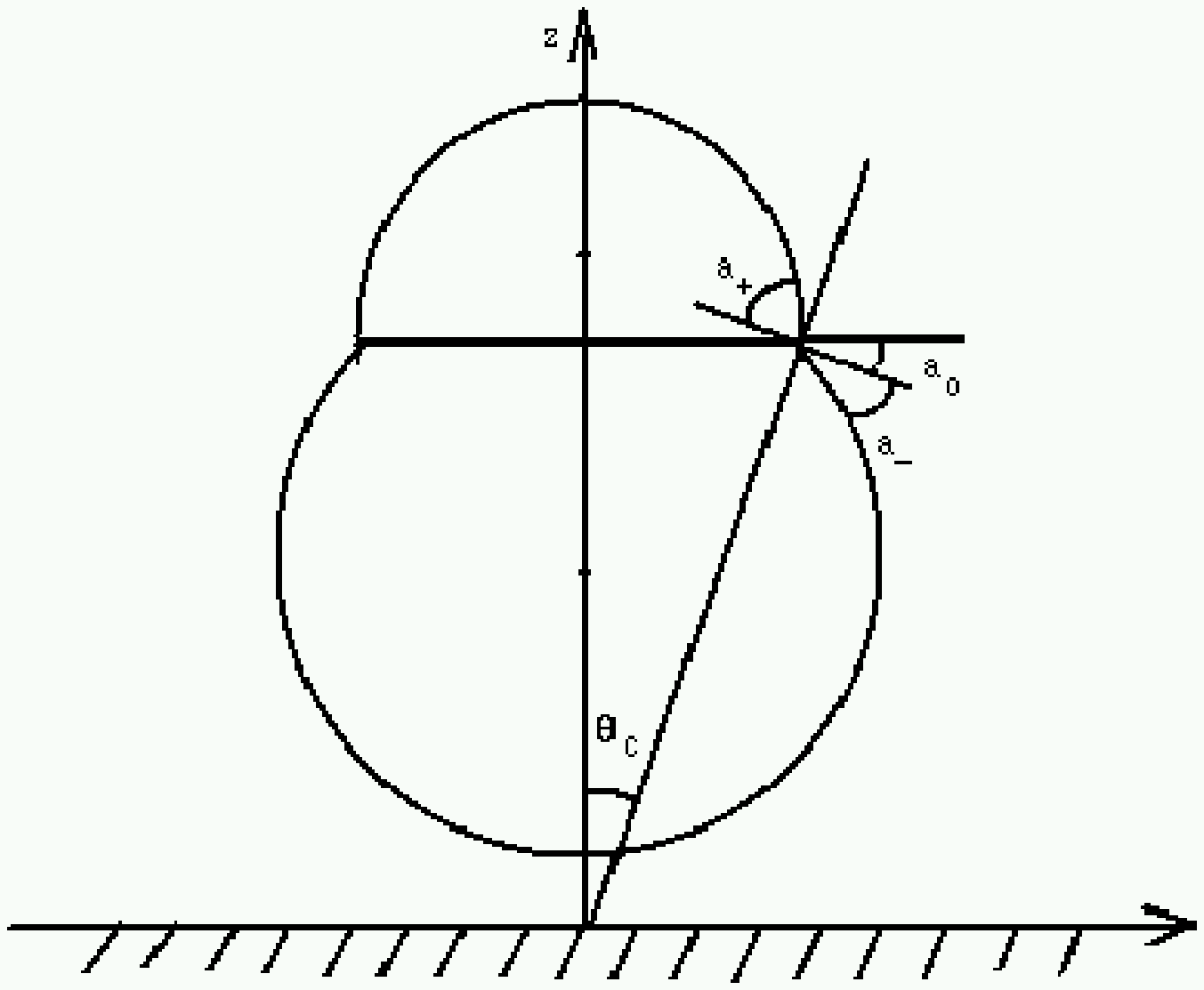}
\end{center}

Apparently, the configuration sketched above has a finite  action
since none of the constituting pieces reaches the AdS boundary ,
hence the tunneling goes with a finite probability which
can be easily computed. Notice that analogously to constructions
in \cite{mms} one needs a brane breaking BPS inequality in order
to have a finite probability of tunneling. An example of this type of brane
in string theory was given in \cite{mms}.

Junction manifold is of the type  $\theta=const$, $\tau=const$,
and since overall rescaling of the solution is not important
(the effective action is invariant under total rescaling,
or, equivalently, under total shift in $\tau$-coordinate),
we take it as follows:
\beq
\label{gum}
\theta=\theta_{c},\; \tau_{c}=0
\eeq
>From Eqs.(\ref{bps2}),(\ref{under2}),(\ref{over2}) one immediately
obtains
\begin{eqnarray}
\label{gum2}
\cos \theta_{c}=z_{0},\nonumber\\
\sinh a = \sinh \tau_{m} \cos\theta_{c}\\
\cosh b = \cosh \rho_{m} \cos\theta_{c}.\nonumber
\end{eqnarray}

The  Eqs.(\ref{bps2}),(\ref{under2}),(\ref{over2}),
(\ref{gum}),(\ref{gum2}) specify the geometry of the
Big Bang bounce. However we still have to define charges
of the branes involved and to verify that the junction conditions are
fulfilled.

The charge conservation, with appropriate choice of orientation
of the branes, reads
\beq
\label{charge}
Q_{0}=Q_{-}-Q_{+}.
\eeq
Hereafter subscripts "0","-", and "+" indicate the  BPS brane,
the undercharged brane and the overcharged brane.
All $Q$'s are assumed to be positive.

The force balance condition obviously reads (we take projections
onto $\partial/\partial\theta$ and onto $\partial/\partial\tau$
directions):
\begin{eqnarray}
\label{balance}
T_{0} \cos\alpha_{0} + T_{-} \cos\alpha_{-}=T_{+}\cos\alpha_{+}
\nonumber\\
T_{0} \sin\alpha_{0} + T_{+} \sin\alpha_{+}=T_{-}\sin\alpha_{-}
\end{eqnarray}
where the angles are defined on fig.5. From geometry of the
picture and using Eqs.(\ref{over2}),(\ref{under2}),(\ref{gum}),
(\ref{gum2}) one obtains
\begin{eqnarray}
\label{angles}
\alpha_{0}=\theta_{c}\nonumber\\
\tan\alpha_{-}=\frac{\sin \theta_{c}\sinh \tau_{m}}{\cosh a}\\
\tan\alpha_{+}=\frac{\sin \theta_{c}\cosh \rho_{m}}{\sinh b}\nonumber
\end{eqnarray}

According to the above definition of $q$ (see Eq.(\ref{action2})) we take the
following parametrization of the tensions of the three pieces of the bounce:
\beq
\label{tension}
T_{0}=Q_{0}T,\;T_{-}=\frac{Q_{-}}{q_{-}}T,\;T_{+}=\frac{Q_{+}}{q_{+}}T
\eeq
where $q_{-}=\tanh \tau_{m}$, $q_{+}=1/\tanh\rho_{m}$.

Substituting Eq.(\ref{tension}) into the force balance condition
Eq.(\ref{balance}) and using the charge
conservation condition Eq.(\ref{charge}) one obtains a linear
system for the charges of R-branes:
\begin{eqnarray}
\label{last}
Q_{-} \left( \cos\theta_{c} + \frac{\cos\alpha_{-}}{q_{-}}\right) -
Q_{+} \left( \cos\theta_{c} + \frac{\cos\alpha_{+}}{q_{+}}\right)=0
\nonumber\\
Q_{-} \left( \sin\theta_{c} - \frac{\sin\alpha_{-}}{q_{-}}\right) -
Q_{+} \left( \sin\theta_{c} - \frac{\sin\alpha_{+}}{q_{+}}\right)=0
\end{eqnarray}
Using Eqs.(\ref{gum2},(\ref{angles}) one can straightforwardly verify
that determinant of this linear
system is equal to zero, so the system is compatible and
defines the ratio of charges of the R-branes at which the
force balance condition Eq.(\ref{balance}) is satisfied.
This completes our construction of the Big Bang bounce.

\section{Conclusion}
Our conclusions are as follows:\\
1. We have described induced brane production in the external field;\\
2. We have found the probability of the  tunneling into the Brane World (a
sort of Big Bang);\\
3. It was shown that there is no need for the negative
tension branes to be included in the scenario with the external field;\\
4. The direct consequence of our approach is the 5d early Universe;\\
5. It was
shown that the simple brane configuration in the $AdS_5$ with the external
field yields the flat 4d Universe.

The work was partially supported by grants CRDF-RP1-2108 , INTAS-97-0103 (K.S)
and INTAS-99-1705 (A.G.)

\end{document}